\documentclass[conference]{IEEEtran}
\IEEEoverridecommandlockouts
\usepackage{cite}
\usepackage{amsmath,amssymb,amsfonts}
\usepackage{algorithmic}
\usepackage{graphicx}
\usepackage{textcomp}
\usepackage{xcolor}
\usepackage{braket}
\usepackage{subcaption}

\def\BibTeX{{\rm B\kern-.05em{\sc i\kern-.025em b}\kern-.08em
    T\kern-.1667em\lower.7ex\hbox{E}\kern-.125emX}}
\begin{document}

\title{Quantum simulation and circuit design for solving multidimensional Poisson equations\\
}

\author{Michael Holzmann$^1$ and Harald K\"ostler$^1$ \\\\
$^1$ Friedrich-Alexander-Universit\"at Erlangen-N\"urnberg, Germany\\
}

\maketitle

\begin{abstract}
Many methods solve Poisson equations by using grid techniques which discretize the problem in each dimension. Most of these algorithms are subject to the curse of dimensionality, so that they need exponential runtime. In the paper "Quantum algorithm and circuit design solving the Poisson equation" a quantum algorithm is shown running in polylog time to produce a quantum state representing the solution of the Poisson equation. In this paper a quantum simulation of an extended circuit design based on this algorithm is made on a classical computer. Our purpose is to test an efficient circuit design which can break the curse of dimensionality on a quantum computer. Due to the exponential rise of the Hilbert space this design is optimized on a small number of qubits. We use Microsoft's Quantum Development Kit and its simulator of an ideal quantum computer to validate the correctness of this algorithm.
\end{abstract}

\section{Introduction}
The goal of this work is to implement a quantum algorithm solving the d-dimensional Poisson equation with Dirichlet boundary conditions:
\begin{align*}
    -\Delta u(x) = f(x) \ , \ \text{x in} \ \Omega \ \ 
    \\
    u(x) = g(x) \ ,  \ \text{x on} \ \delta\Omega \ \ \Omega = (0,1)^d
\end{align*}
One way to solve this problem is to discretize $\Omega$ in M+1 grid points in each dimension. M is an exponent of base 2 in this work. The solution u(x) is a vector of $(\text{M}-1)^d$ entries. To calculate a problem of one dimension a linear equation system has to be solved:

\begin{align*}
\resizebox{1.0\hsize}{!}{
   $\frac{1}{h^2} A_{d=1} \begin{pmatrix}u_1 \\ \\ \vdots \\ \\ u_{m-1}\end{pmatrix}=  \frac{1}{h^2} \begin{pmatrix}2&-1 & & 0\\-1& \ddots &\ddots \\ 0& \ddots& \ddots \\ & &  -1& 2\end{pmatrix} \begin{pmatrix}u_1 \\ \\ \vdots \\ \\ u_{m-1}\end{pmatrix} = \begin{pmatrix}f_1 + \frac{1}{h^2}u_0\\ \\ \vdots \\ \\ f_m+\frac{1}{h^2}u_m\end{pmatrix} =\Vec{b}$}
\end{align*}
For multidimensional problems the linear equation uses matrix $A_{d}$, which can described as \cite{refer}:

\begin{align}
     A_d = \sum_{i=1}^d {\bigotimes^{i-1} I \otimes A_{d=1} \otimes \bigotimes^{d-i} I }
     \label{eq1}
\end{align}

Matrix $A_{d}$ has the dimension of $(\text{M}-1)^d \times(\text{M}-1)^d$. The best numerical algorithms for solving this problem run polynomially to matrix size\cite{cg}, so the runtime increases exponentially with the dimension of the problem. In this paper a quantum algorithm is used to produce a quantum state representing the normalized solution of the problem. Since this technique runs in polylog time the curse of dimensionality can be broken.
\section{Quantum algorithm and circuit design}
The quantum algorithm for solving linear equations is based on the HHL09 algorithm\cite{hhl}. If successfull this algorithm replaces the input $\Vec{b}$ with the normalized solution $\Vec{u}$ of $u(x)$. The algorithm used in this work follows several steps:
\begin{itemize}
    \item Produce a quantum state $\sum_{j}\beta_j \ket{j}$ in RegC. $ \beta_j \in \Vec{b}$
    \item Use Phase Estimation Algorithm (PEA) on RegC and RegB. Register B consists of n qubits. In the process of PEA several Hamiltonian simulations of U = $e^{ \frac{1}{h^2}A_d t }$ with t = $2\pi i \frac{1}{2^n} 2^k$ \,$k=0,..., n-1$ are applied to RegC. PEA entangles the eigenvalues $\lambda_j$ of $A_d$ in RegB with the eigenstates $\Vec{u_j}$ in RegC: $\sum_{j} b_j \ket{\approx \lambda_j} \ket{\Vec{u_j}}$ with $b_j = \braket{\Vec{b}|\Vec{u_j}}$
    \item Calculate the reciprocal of the eigenvalues in RegA. The system has now the state: $\sum_{j} b_j  \ket{\frac{1}{k_j}}\ket{k_j} \ket{\Vec{u_j}}$ with $k_j \approx \lambda_j$
    \item Apply a controlled rotation on an ancilla qubit to produce following system: $\sum_{j} b_j  \ket{\frac{1}{k_j}}\ket{k_j} \ket{\Vec{u_j}} (\sqrt{1-\frac{\alpha^2}{k_j^2}}) \ket{0} + \frac{\alpha}{k_j}\ket{1})$ with $k_j \approx \lambda_j$ and amplitude factor $\alpha$
    \item Uncompute RegA and RegB 
    \item Measure the ancilla qubit. If the measurement of the qubit results in state $\ket{1}$, the algorithm successfully transforms RegC into the normalized solution $\Vec{u}$ = $C\sum_{j} b_j  \frac{\alpha}{k_j} \ket{\Vec{u_j}}  =  C A^{-1}\Vec{b}$ with C as a normalization constant. Otherwise the algorithm has to be restarted.
    
\end{itemize}

\subsection{Encoding of register C}
For the first step the right side of the linear equation has to be encoded in register C. If $\beta_j$ can be calculated efficiently, a quantum oracle can be applied to a zero state register to produce $\sum_{j}\beta_j \ket{j}$ in polylog time \cite{stateprep}. Since $\log_2 (\text{M})$ qubits can encode M states, the zero state is not used and has amplitude zero in the one-dimensional case. For multidimensional problems $d \log_2 (\text{M})$ qubits are needed. To solve these problems Hamiltonian simulations of $A_{d=1}$ each acting on $\log_2 (\text{M})$ qubits are used. Because every Hamiltonian simulation does not allow a zero state, every $\log_2(\text{M})$ qubit block of register C must not be in zero state. This yields $\text{M}^d - (\text{M}-1)^d$ invalid states. i.e for $d=1$ and $\text{M}=4$ $\beta_1$ is the amplitude of state $\ket{1}$, while for the two-dimensional case it is encoded in the amplitude of state $\ket{5}$. Since register C is used for the input and output of the algorithm, the same encoding applies for the normalized solution.

\subsection{Phase Estimation Algorithm of $A_d$ (PEA)}
\begin{figure}[h]
\centerline{\includegraphics[width=7cm]{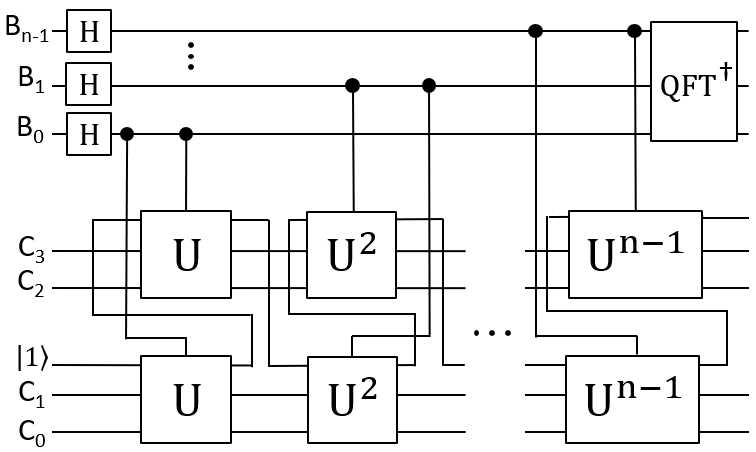}}
\caption{Circuit for PEA with $\text{M}=4$ and $\text{d}=2$}
\label{pea}
\end{figure}

One subroutine of the algorithm is the one-dimensional Hamiltonian simulation $U = e^{\frac{1}{h^2}A_1 t}$. Using the spectral theorem \cite{refer}:
\begin{align}
   e^{\frac{1}{h^2}A_1 t} =  S e^{t\Delta} S \,\,\, ,\ \Delta = \begin{pmatrix}\lambda_1&0&0&0\\0&\lambda_2&0&0\\0&0&\ddots&0\\0&0&0&\lambda_{M-1}\end{pmatrix}
\end{align}
S desribes the discrete sine transform acting on $\log_2(\text{M})+1$ qubits, which can be implemented with the quantum fourier transform (QFT), two transformation circuits and one ancilla qubit \cite{sine}. For the Hamiltonian simulation of a 1-sparse matrix $e^{t\Delta}$ two oracle calls are needed. The first one calculates the eigenvalues $\lambda_j$ of the one-dimensional Poisson matrix, which are used to produce the right rotation in the secondary oracle call \cite{hamiltonian}. Since $A_1$ is a toeplitz matrix, $\lambda_j$ are \cite{eigenval}:
\begin{align}
     \lambda_j = 4\text{M}^2\sin^2(\frac{j\pi}{2\text{M}}) \ \, \, \, \ j=1, ..., \text{M}-1 \label{eigenvalues}
\end{align}
The sine function can be approximated with a quantum oracle using repeated squaring \cite{refer}. Even though it can be well calculated with respect to M, quantum arithmetics use many ancilla qubits to run. Since the referenced quantum algorithm of the repeated squaring process results in multiple ancilla registers, the Hilbert space rises exponentially. In the paper ``Quantum Fast Poisson Solver: the algorithm and modular circuit design''\cite{refer2} a module is shown calculating these eigenvalues with $O(n \text{M})$ qubits where n describes the size of register B. Since n depends on the matrix size M their simulations need to run on a supercomputer. We instead use a quantum circuit design which runs linear to M and can be simulated easily on a classical computer. This gives us the possiblity to simulate multidimensional problems. To entangle the eigenvalues with their eigenstates PEA needs to run several Hamiltonian simulations with the power up to $2^{n-1}$. 
\begin{align}
    U^{2^k} = S e^{t\Delta2^k} S \, \,\, \text{for} \,\, k=0, \cdots, n-1 
\end{align} Thus the application of multiple U can be directly encoded in the phase shifts of $e^{t\Delta}$, so that the runtime of $U^{2^k}$ does not increase. The Hamiltonian simulation of a diagonal matrix can be split into:
\begin{align}
\resizebox{0.44\textwidth}{!}{%
$e^{t\Delta2^k} = \prod_{j=1}^{M-1} diag(1_0, ..., 1_{j-1}, e^{\lambda_jt2^k}, 1_{j+1}, ..., 1_{M-1})
$}
\end{align}
This can be implemented with $\text{M}-1 \ \text{phase shift gates} \ R_j = \begin{pmatrix}1 & 0\\0 & e^{i\lambda_jt2^k}\end{pmatrix}$. Because we do not use any oracle producing these phases, they have to be calculated classicly. This is the reason why this typ of circuit design does not give any exponential speedup in the one-dimensional case. 
\begin{figure}[t]
\centerline{\includegraphics[width=9cm]{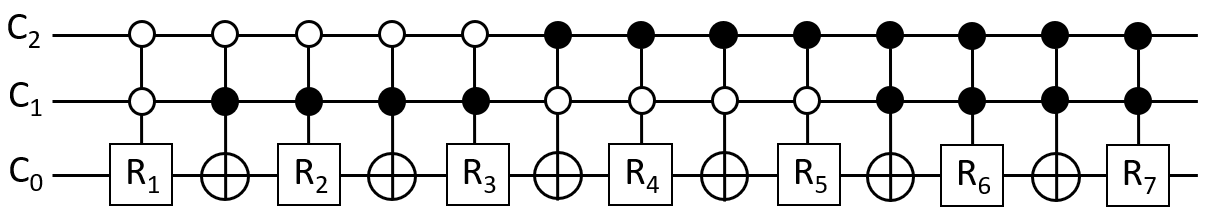}}
\caption{Hamiltonian simulation of a 1-sparse matrix with seven eigenvalues using $\text{M}-1$ phase shift gates}
%\label{fig}
\end{figure}
It can be demonstrated how to simulate multidimensional Poisson matrices by using U. The splitting formula (\ref{eq1}) and its Hamiltonian simulation show that the simulation of a d-dimensional problem can be implemented with d one-dimensional Hamiltonian simulations \cite{refer}. In fact this breaks the curse of dimensionality and brings the exponential quantum speed up with respect to d.
Figure (\ref{pea}) shows the quantum circuit for PEA of a two-dimensional problem.
The ancilla qubit, which is needed for applying the discrete sine transform, can be reused for every Hamiltonian simulation, because it is kept in state $\ket{1}$. This circuit design is the reason for the specific encoding of register C, since every $\log_2(\text{M})$ qubit block uses the subroutine U separately. The phase estimation algorithm entangles eigenstates with eigenvalues of $A_d$ and produces the following system:
\begin{align*}
    \ket{\psi} = \sum_{j} b_j \ket{\approx \lambda_j} \ket{\Vec{u_j}} \text{ with }\, b_j = \braket{\Vec{b}|\Vec{u_j}}
\end{align*}
\subsection{Reciprocal calculation of $\lambda_j$ (INV)}
For the next step the eigenvalues have to be entangled with its reciprocals. We use one iteration of the Newton-Raphson-Division method to calculate their values:
\begin{align}
    \lambda^{-1} \approx 2x_0 -\lambda x_0^2
\end{align}
For the start value $x_0$ we use an approximation of the reciprocal:
\begin{align}
    x_0 = \frac{1}{2^p} , \ p \in \mathbb{R} \ \text{with}\ |2^p-\lambda_j|_{min}
\end{align}
Let $k=k_0, ..., k_{n-1}$ be the binary representation of $\lambda_j$ as an integer. The following circuit describes a method to store $x_0$ in a new n-sized register A for $\lambda_j \geq 2$. The circuit iterates through all qubits in register B starting with the qubit with the highest value. If it reads a $\ket{11}$ block, $x_0$ will be rounded off, otherwise $x_0$ is greater than the reciprocal of $\lambda_j$. The iteration stops after the first one is read. The ancilla qubit indicates at the end whether the inversion happened or not. 
\begin{figure}[h]
\centerline{\includegraphics[width=9cm]{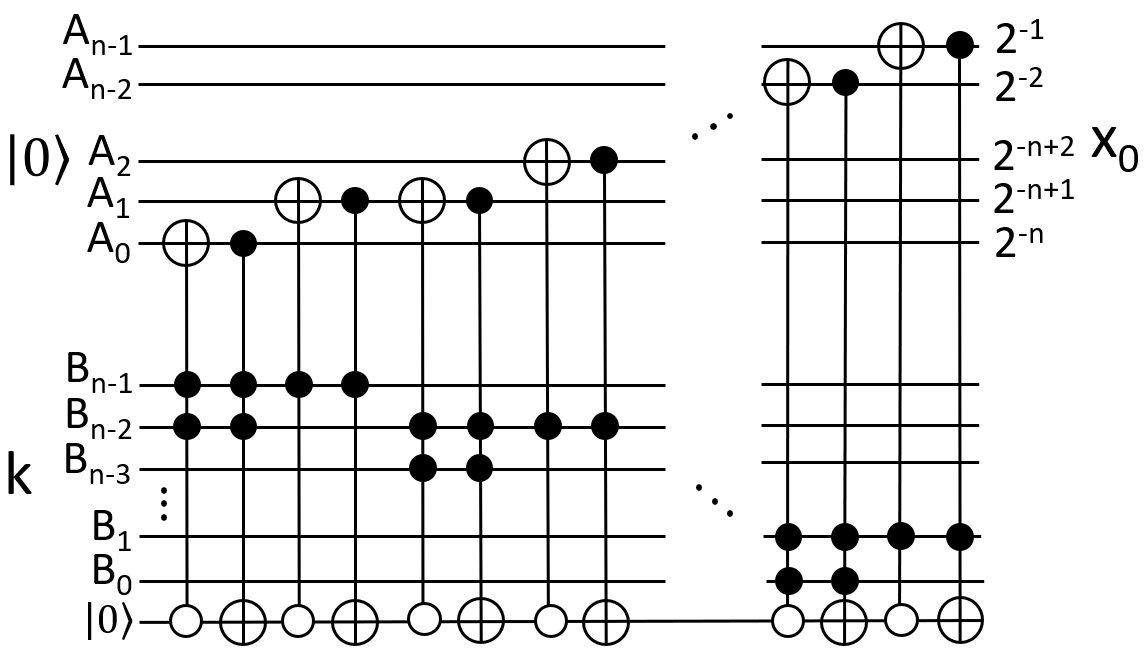}}
\caption{Quantum circuit to transform $\ket{0}\ket{k_j}$ into $\ket{x_{0_j}}\ket{k_j}$}
%\label{fig}
\end{figure}
To get a more accurate reciprocal of $\lambda_j$ the first iteration of the newton method gets calculated:
\begin{align}
    x_1= 2^{-p+1} - \lambda 2^{-2p}
\end{align}
This could be implemented with one quantum subtraction, one multiplication and one squaring circuit. Most of the known quantum multiplication circuits store the result in an additional two times sized ancilla register \cite{arithmetic}. Even if truncation of the result is used, the simulation cost of the system would increase rapidly with respect to n. In our work the calculation of $x_1$ is performed in the rotational part of the algorithm. Register A and register B together contain the whole information about $x_1$. Instead calculating $x_1$ as a fixed precision number, our inversion modul produces a floating point number, where RegB can be unterstood as the mantissa and RegA as the binary exponent. The quantum state can now be described as:
\begin{align*}
    INV\ket{\psi} = \sum_{j} b_j  \ket{\Vec{u_j}} \ket{k_j} \ket{k_j^{-1}} \text{ with }\, k_j^{-1} = x_{0_j}
\end{align*}
\subsection{Rotation of the reciprocal (ROT)}
The goal of this module is to rotate the $\alpha$-fold reciprocal $x_1$ into the amplitude of an ancilla qubit. Let $\phi_0, ..., \phi_{n-1} \ \phi_i \in (0,1)$ be the binary representation of a number with the value order $2^{-1}, ..., 2^{-n}$. The rotaton of a n-qubit large number can be implemented with n controlled rotations about the  y-axes of the Bloch sphere \cite{refer}.
\begin{equation}
\begin{gathered}
    R_y(\alpha \phi)= \phi_0 R_y(\alpha 2^{-1}) \cdot \phi_1 R_y(\alpha 2^{-2}) \cdot ... \cdot \phi_{n-1} R_y(\alpha 2^{-n}) \label{rot}
\\
    \text{with } R_y(\alpha \phi ) = \begin{pmatrix} \cos \alpha \phi  & -\sin \alpha \phi \\ \sin \alpha \phi & \cos \alpha \phi \end{pmatrix}
\end{gathered}
\end{equation}
Every scalar factor of the angle $\phi$ can be implemented directly in the rotation gates. To rotate the ancilla qubit about the angle $\alpha x_1$ we need two rotations. The first part applies the angle $\alpha 2^{-p+1}$. Since $x_0$ is not known before runtime, we have n different cases. This can be solved with n different controlled rotations. The second part rotates the floating point number $-\alpha \lambda 2^{-2p}$. The mantissa $\alpha \lambda$ also can be implemented with n different controlled rotations described in formula (\ref{rot}). The multiplication with the binary exponent could be done by shifting the mantissa. This yields bigger registers and larger simulation costs. We instead rotate the mantissa n times shifted by the exponential factor each one belongs. Let $a_0, ..., a_{n-1} \ a_i \in (0,1)$ be the binary representation of $x_0$. The rotation of the floating point number $\alpha \lambda 2^{-2p}$ can be split into:
\begin{align}
    R_y(\alpha \lambda 2^{-2p} ) = \prod_{i=0}^{n-1} a_i R_y(\alpha \lambda 2^{-2(i+1)}) 
\end{align}
The subtraction itself is implemented by rotating in the other direction.
This method keeps the number of qubits small and allows us to simulate more difficult Poisson problems, however it needs much more runtime ($O(n^2)$ rotation gates). The quantum state of the system after using this module looks like:
\begin{align}
  \sum_{j} b_j  \ket{\Vec{u_j}} \ket{k_j} \ket{k_j^{-1}} (\sqrt{(1-\sin^2(\frac{\alpha}{k_j}))}\ket{0} + \sin(\frac{\alpha}{k_j})\ket{1}) 
\end{align}
Since the amplitude $\sin(\frac{\alpha}{k_j})$ is not a multiple of the reciprocal, we have to adapt the rotation angle to $\arcsin(\frac{\alpha}{k_j})$. The referenced paper shows an algorithm for approximating this by using bisection and their sine function module\cite{refer}. Because this method uses repeated squaring as a subroutine in the sine function, the simulation cost of the system would increase exponentially for the same reason as the eigenvalue calculation. Thus we skip this part of the algorithm and use the small angle approximation of the sine function: $\sin(\frac{\alpha}{k_j}) \approx \frac{\alpha}{k_j} $ for $ \frac{\alpha}{k_j} \in [0;0.5]$. Skipping the arcsin module gives us the opportunity to shift the calculation of $x_1$ in the rotational part of the algorithm. Thus we do not need any quantum arithmetic circuits.
\subsection{Amplitude factor $\alpha$}
After uncomputing registers A and B a measurement of the rotated ancilla qubit leads with success probability $\Omega_{succ}$ to state $\ket{1}$.
\begin{align}
    \Omega_{succ}= \sum_{j} b_j^2 \sin^2(\frac{\alpha}{\approx \lambda_j})
\end{align}
If the measurement leads to one, register C collapses to the normalized solution $\Vec{u}$ of the problem:
\begin{align*}
    \Vec{u} = C\sum_{j} b_j  \frac{\alpha}{\approx \lambda_j} \ket{\Vec{u_j}}  \text{ with } C=(\sqrt{\sum_j (\frac{b_j\alpha}{\approx \lambda_j})^2})^{-1}
\end{align*}
Due to the normalization of C small amplitude factors do not influence the solution, but they do boost the success probability of the algorithm. Doubling the amplitude of the qubit means four times better success rates. Because we do not want to produce big errors, $\alpha$ has a upper boundary of $\frac{\lambda_1}{2}$. Thus the amplitude for state $\ket{1}$ is at least $\frac{1}{2 \kappa}$, so that the runtime of the algorithm is $O(4\kappa^2)$. By using an arcsin module $\alpha$ can be doubled, which results in the same runtime described in the HHL09 algorithm. Other methods like variable time amplitude amplification may be used to achieve closely linear runtimes to $\kappa$\cite{amplitude_ampl}.
\begin{figure*}[t]
\centerline{\includegraphics[width=16cm]{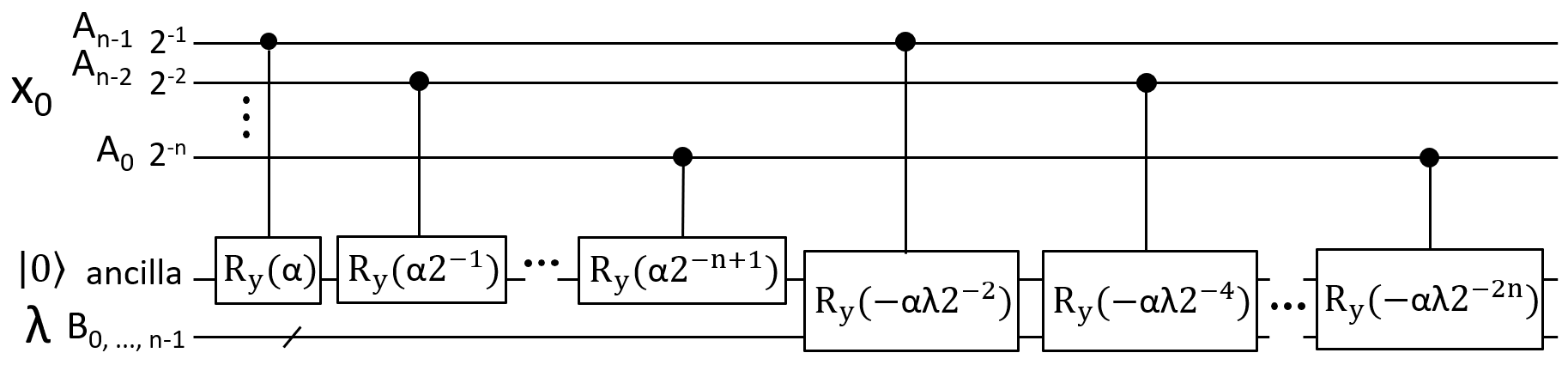}}
\caption{Rotation circuit to transform state $\ket{\lambda_j}\ket{x_{0_j}}\ket{0}$ into $\ket{\lambda_j}\ket{x_{0_j}}(\sqrt{1-\sin^2(\alpha x_{1_j})} \ket{0} + \sin(\alpha x_{1_j}) \ket{1})$}
%\label{fig}
\end{figure*}
\section{Results}
To validate our circuit design for the algorithm we use following example for a problem with M = 4 and d = 1:
\begin{align*}
   \text{Ex. 1:} \, \, A_1= 16 \cdot \begin{pmatrix}2 & -1 & 0 \\ -1 & 2 & -1 \\ 0 & -1 &2 \end{pmatrix} \ \ \Vec{b} = \begin{pmatrix}1 \\ 0 \\ 0 \end{pmatrix}
\end{align*}
with eigenvalues $\lambda_1 = 9.37, \ \lambda_2 = 32, \ \lambda_3 = 54.62 $, eigenvectors:
\begin{align*}
    \Vec{u_1} = \begin{pmatrix}0.5 \\ \frac{\sqrt{2}}{2} \\ 0.5 \end{pmatrix}, \
    \Vec{u_2} = \begin{pmatrix}\frac{1}{\sqrt{2}} \\ 0 \\ -\frac{1}{\sqrt{2}} \end{pmatrix}, \
    \Vec{u_3} = \begin{pmatrix}0.5 \\ -\frac{\sqrt{2}}{2} \\ 0.5 \end{pmatrix}
\end{align*}
and $b_1=\frac{1}{2},\ b_2=\frac{1}{\sqrt{2}},\ b_3=\frac{1}{2}$.
First we check if the phase estimation algorithm of the Hamiltonian simulation produces the right quantum state $\ket{\psi}$. For the register size of RegA and RegB we choose that $\lambda_j$ can be quantified as fixed precision integer. Based on formula (\ref{eigenvalues})  we can estimate a upper boundary for the eigenvalues:
\begin{align}
    \lambda_{max} = 4d\text{M}^2\sin^2(\frac{(\text{M}-1)\pi}{2\text{M}}) \leq 4d\text{M}^2
\end{align}
Thus we use $n= 2 + \log_2d + 2\log_2(\text{M})$ qubits for registers A and B. One way to varify the achieved quantum state $\ket{\psi}$ is to measure register B. Consequently the wave function collapses with probability $b_j^2$ to $\ket{\lambda_j}\ket{\Vec{u}_j}$.
\begin{figure}[h]
\advance\leftskip-0.25cm
\includegraphics[width=9cm]{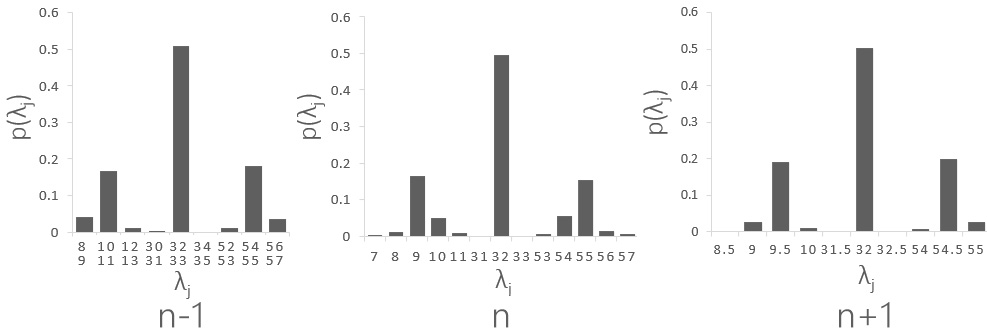}
\caption{Eigenvalue destribution for 2000 measurements with register size n-1, n and n+1}
%\label{fig}
\end{figure}
Measurements show the expected result of PEA. The state for the highest eigenvalue is exactly represented with its probability. Since eigenvalue $\lambda_1$ and $\lambda_3$ can not be stored as an integer, PEA builds a distribution around its eigenvalue. The probability to get the nearest eigenvalue approximation is at least $\frac{4}{\pi^2}$ \cite{pea}. We can use more qubits to increase the resolution which leads to a more precise eigenvalue distribution. It is possible to reduce the register size to $\text{n} - \text{k}$, where $2^k$ is the best smallest eigenvalue approximation. This leads to a resolution error of $2^{k-1}$. Thus k less qubits are needed and the register size is $O(\kappa)$. We can use this method to reduce the simulation cost of the system. By utilizing the register dump method of Microsoft's quantum simulator we can check the state of register C. By subtracting the global phase we can see, that the amplitudes and phases correspond to the right eigenstate which belongs to its eigenvalue.
\\\\
In the following we show the influence of the amplitude factor $\alpha$ to the solution and the success probability of the algorithm. The algorithm transforms register C with probability of success $\Omega_{succ}$ into a state of the form $\sum_x u(x) \ket{x}$. We use Microsoft's quantum simulator to read out the amplitudes of register C. The output encoding for register C has to be considered. The success probability depends on $b_j$, $\alpha$ and the condition number of $A_d$. By increasing $\alpha$ we can rotate a multiple of the reciprocal. For a valid small angle approximation doubling $\alpha$ results in a four times higher $\Omega_{succ}$.
\begin{figure}[h]
\center
\noindent

\begin{subfigure}{.23\textwidth}

  \includegraphics[width=4cm]{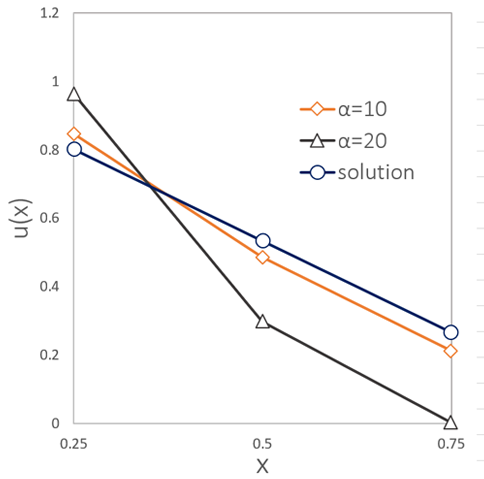}

\end{subfigure}%
\begin{subfigure}{.23\textwidth}

  \includegraphics[width=3.8cm]{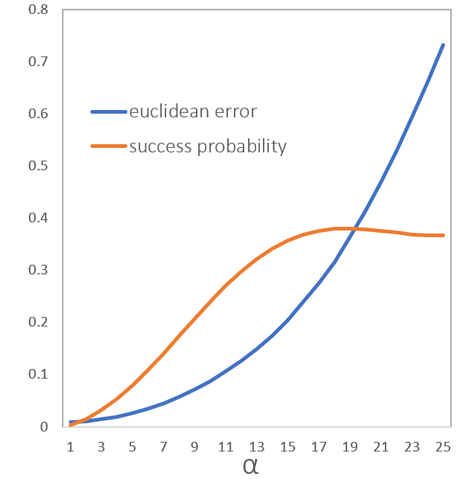}
  \
\end{subfigure}

\caption{Measurements of register C for example one with different $\alpha$}
\label{m4d1}
\end{figure}
By choosing $\alpha$ above the upper boundary $\frac{\lambda_1}{2}$ on the one hand we violate the small angle approximation, on the other the amplitude may be over rotated. This is the reason why high amplitude factors produce high errors in the output. If $b_1$ is close to zero, we can set the upper boundary for $\alpha$ to the next higher eigenvalue $\frac{\lambda_2}{2}$ , since the error of $\sin(\frac{\alpha}{\lambda_1})$ is multiplied with $b_1$. This concept may be continued to higher upper boundaries. Because $b_j$ is not known, one method could be to increase $\alpha$ stepwise until the algorithm shows success with a certain precision.
\\\\
It is posssible to solve more complex problems with higher dimensions. The described algorithm solves multidimensional problems by simulating multiple Hamiltonian simulations in the PEA step. When we increase the dimension of the problem by one, at least $\log_2(\text{M})$ additional qubits are needed to run, which leads to a linear qubit count to d. All together we use $7+2\log_2d+(4+d)\log_2(\text{M})$ qubits. i.e. for $\text{M}=8$ and $d=2$ 27 qubits are needed.
\begin{figure}[t]
\leftskip-0.7cm
  \noindent

\begin{subfigure}{0.18\textwidth}
\leftskip0.3cm
  \includegraphics[width=3.2cm]{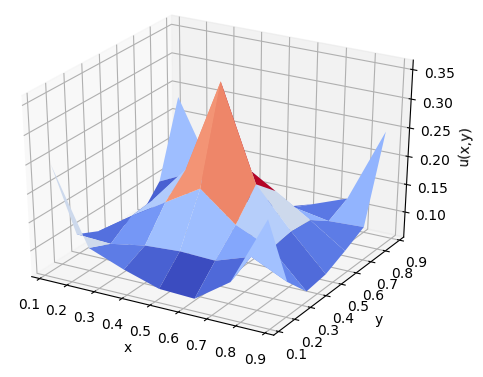}
  \caption{solution}
  \label{fig:sub1}
\end{subfigure}%
\begin{subfigure}{.18\textwidth}
\leftskip0.2cm
  \includegraphics[width=3.2cm]{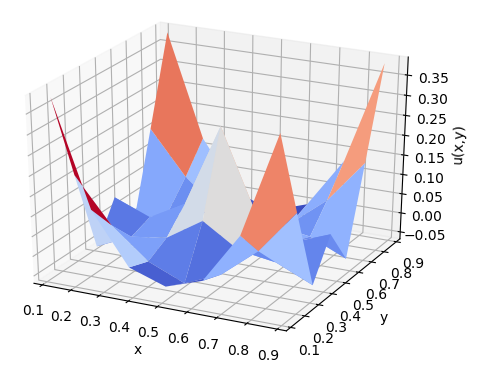}
  \caption{$\alpha=30, \ \Omega \approx 3\%$}
  %\label{fig:sub2}
\end{subfigure}%
\begin{subfigure}{.18\textwidth}
\leftskip0.2cm
  \includegraphics[width=3.2cm]{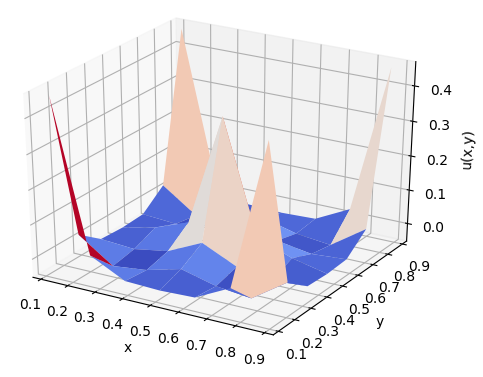}
  \caption{$\alpha=300, \ \Omega \approx 71\%$}
  %\label{fig:sub2}
\end{subfigure}
\caption{Measured output for different $\alpha$ with $\text{M}=8$ and $\text{d}=2$}
\label{m8d2}
\end{figure}
 Figure (\ref{m8d2}) shows the measured output for a two dimensional problem. By changing $\alpha$ we can choose between accuracy and success probability. While (b) violates the small angle approximation, (c) overrotates its amplitude in addition. This results in higher errors, but better success rates. Thus this circuit design can be used to achieve an approximation of the normalized solution with high probability. To get more accurate results the algorithm has to repeated $O(4\kappa^2)$ with an amplitude factor less than $\frac{\lambda_1}{2}$. Amplitude amplification techniques may be used to get better runtimes to $\kappa$.
 \\\\
 Since the values of $\Vec{u}$ are encoded in the amplitudes of register C, we can not read them out directly. To get the full solution of the equation the algorithm has to be successfully repeated at least $O((\text{M}-1)^d)$ times. This does not give us any quantum advantage over classical methods, though it does if one is interested in an expectation value of an operator acting on $\Vec{u}$. For example a measurement of register C may collapse with high probability into a state $\ket{x}$ which represents the position of a maximum in the solution of the problem. Thus it is possible to get some correlated information about the solution by using several quantum operators. \\\\
 We showed a modular circuit design which can solve multidimensional Poisson equations and can be simulated on a simple classical computer. Since we only use a small number of qubits the memory consumption is kept low. It may also be possible to test this design on a real quantum computer. By adding some error correcting qubits the today's most advanced quantum computers could run simple problems with a similar circuit design.

\vspace{12pt}


\begin{thebibliography}{00}
\bibitem{refer} Y. Cao, A. Papageorgiou, I. Petras, J. Traub and Sabre Kais ``Quantum algorithm and circuit design solving the
Poisson equation'', 2012.
\bibitem{refer2} S. Wang, Z. Wang, W. Li, L. Fan, Z. Wei and Yongjian Gu ``Quantum Fast Poisson Solver: the algorithm and modular circuit design '', 2019.
\bibitem{cg} J. Shewchuk, ''An introduction to the conjugate gradient method without the agonizing pain'', 1994.
\bibitem{hhl} A. W. Harrow, A. Hassidim and S. Lloyd, ''Quantum algorithm for solving linear
systems of equations'', 2008.
\bibitem{stateprep} A. N. Soklakov and R. Schack, ''Efficient state preparation for a register of quantumbits'', 2004.
\bibitem{sine} A. Klappenecker and M. Roetteler, ''Discrete cosine transforms on quantum computers'', 2001
\bibitem{hamiltonian} A. M. Childs, ''Quantum Information Processing in Continuous Time'', 2004.
\bibitem{eigenval} D. Kulkarni, D. Schmidt, and S.-K. Tsui, ''Eigenvalues of tridiagonal pseudotoeplitz matrices'', 1997
\bibitem{arithmetic} V. Vedral, A. Barenco, and A. Ekert, ''Quantum networks for elementary arithmetic operations'', 1995.

\bibitem{amplitude_ampl} A. Ambainis, ''Variable time amplitude amplification and a faster quantum
algorithm for solving systems of linear equations'', 2010.

\bibitem{pea}R. Cleve, A. Ekert, C. Macchiavello and M. Mosca, ''Quantum Algorithms Revisited'', 1997
\end{thebibliography}
\end{document}